# Fluctuations, Higher Order Anharmonicities, and Landau Expansion for Barium Titanate

A. I. Sokolov

*St. Petersburg Electrotechnical University, Professor Popov street 5, St. Petersburg, 197376 Russia*
*e-mail: ais2002@mail.ru*

**Abstract**—The correct description of the ferroelectric phase transition in barium titanate requires that, in the expansion of the free energy, not only the sixth-order terms but also the eighth-order terms should be retained. One more unusual feature of the BaTiO$_3$ compound is that the coefficients $B_1$ and $B_2$ of the terms $P_x^4$ and $P_x^2 P_y^2$ in the Landau expansion depend on the temperature. It is demonstrated that the temperature dependence of the coefficients $B_1$ and $B_2$ can be associated with the thermal fluctuations of polarization under conditions where the fourth-order anharmonic constants are anomalously small; i.e., a nonlinearity of the type $P^4$ and higher order anharmonicities make comparable contributions. Regular (noncritical) fluctuation contributions to the coefficients $B_1$ and $B_2$ are calculated to the first order in the sixth- and eighth-order anharmonic constants. Both contributions increase monotonically with increasing temperature, which is in agreement with the experimental dependences of these coefficients. The proposed theory without additional assumptions makes it possible to determine the ratio of the fluctuation components of the coefficients $B_1$ and $B_2$. This ratio also appears to be very close to that observed in the experiment.

The phenomenological theory was first successfully applied to the description of phase transitions in barium titanate nearly 60 years ago [1, 2]. As experimental data have accumulated, it has been revealed, however, that the temperature dependences of the spontaneous polarization and the permittivity can be reproduced and the structure of the phase diagram of the BaTiO$_3$ compound as a whole can be reconstructed only under the assumption that the coefficients $B_1$ and $B_2$ of the terms $P_x^4$ and $P_x^2 P_y^2$ in the Landau–Devonshire expansion depend on the temperature [3–7]. This dependence is so strong that the fourth-order coefficients change signs at temperatures that differ from the ferroelectric phase transition temperature ($T_0 \approx 400$ K) by only 40–70 K.

In recent years, it has been established that the Landau expansion for the BaTiO$_3$ compound exhibits one more unusual property. It turned out that, in order to correctly describe the ferroelectric phase transitions in this material, it is necessary to retain not only the sixth-order terms but also the eighth-order terms in the expansion of the thermodynamic potential [8–12].

Initially, one of the motivations for including terms of the $P^8$ type was to eliminate the strong temperature dependence of the coefficients $B_1$ and $B_2$ [10]. However, more recently, it has been established that, within this approach, the above problem has defied solution [11, 12].

In this paper, we will demonstrate that both of the aforementioned features of the Landau expansion for barium titanate can have the same origin; namely, they can be a consequence of the anomalously strong anharmonicity of the ferroelectric subsystem of the crystal. An anomalously strong anharmonicity is considered to mean the situation where terms of the $\beta\varphi^4$, $\gamma\varphi^6$, and $\delta\varphi^8$ types (here, $\varphi$ is the normal coordinate corresponding to the soft mode) in the lattice Hamiltonian make comparable contributions to the thermodynamics of the ferroelectric. It is obvious that this situation arises when, for some reason, the constants $\beta$ and $\gamma$ are numerically small. In this case, their analogs in the Landau expansion, i.e., the coefficients $B$ and $\Gamma$ of the terms $P^4$ and $P^6$, can differ significantly from their bare values $\beta$ and $\gamma$ due to the contributions determined by the thermal fluctuations of polarization. By the fluctuation contributions are meant here not the corrections to the results of the Landau theory that diverge at $T \longrightarrow T_c$ and which, as a rule, are kept in mind in the study of fluctuation effects (see, for example, [5, 13–15]). Here, we are dealing with the regular (noncritical) fluctuation components of the coefficients in the Landau expansion, which increase monotonically with increasing

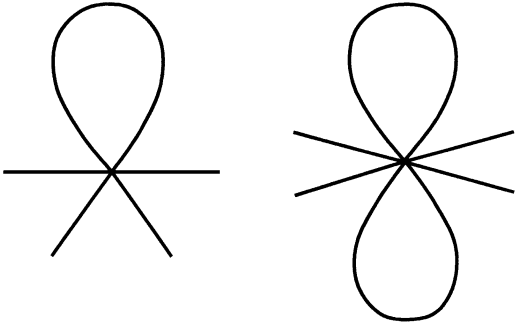

Feynman diagrams corresponding to the first-order corrections to the coefficients $B_1$ and $B_2$.

temperature and have no pronounced features at the phase transition points.

Let us determine the fluctuation corrections to the coefficients $B_1$ and $B_2$ in the Landau expansion to the first order in the sixth- and eighth-order anharmonic constants. As the initial Hamiltonian, we consider a model Hamiltonian for a cubic ferroelectric that contains all invariants allowed by the symmetry of the problem. The anharmonic part of this Hamiltonian has the form

$$H_{anh} = \frac{\beta_1}{4!}(\varphi_1^4 + \varphi_2^4 + \varphi_3^4)$$

$$+ \frac{\beta_2}{4}(\varphi_1^2\varphi_2^2 + \varphi_2^2\varphi_3^2 + \varphi_3^2\varphi_1^2) + \frac{\gamma_1}{6!}(\varphi_1^6 + \varphi_2^6 + \varphi_3^6)$$

$$+ \frac{\gamma_2}{48}[\varphi_1^2(\varphi_2^4 + \varphi_3^4) + \varphi_2^2(\varphi_3^4 + \varphi_1^4) + \varphi_3^2(\varphi_1^4 + \varphi_2^4)]$$

$$+ \frac{\gamma_3}{8}\varphi_1^2\varphi_2^2\varphi_3^2 + \frac{\delta_1}{8!}(\varphi_1^8 + \varphi_2^8 + \varphi_3^8) \quad (1)$$

$$+ \frac{\delta_2}{6!2!}[\varphi_1^6(\varphi_2^2 + \varphi_3^2) + \varphi_2^6(\varphi_3^2 + \varphi_1^2) + \varphi_3^6(\varphi_1^2 + \varphi_2^2)]$$

$$+ \frac{\delta_3}{(4!)^2}(\varphi_1^4\varphi_2^4 + \varphi_2^4\varphi_3^4 + \varphi_3^4\varphi_1^4)$$

$$+ \frac{\delta_4}{96}(\varphi_1^4\varphi_2^2\varphi_3^2 + \varphi_1^2\varphi_2^4\varphi_3^2 + \varphi_1^2\varphi_2^2\varphi_3^4).$$

Here, $\varphi = (\varphi_1, \varphi_2, \varphi_3)$ is the vector field describing the dynamics of the ferroelectric subsystem and the numerical coefficients are of combinatorial origin; i.e., they are chosen from the convenience of the calculation of the Feynman diagrams. As is known, the coefficient of the term $P^{2k}$ in the expansion of the free energy in powers of the order parameter is equal to the total 1-irreducible vertex with $2k$ external lines $\Gamma_{2k}(\mathbf{q}_1, \mathbf{q}_2, \ldots, \mathbf{q}_{2k})$, which is taken on zero momenta (see, for example, [16–18]). Therefore, in our case, we should determine two four-tail vertices that have the same tensor structure as the invariants corresponding to the coefficients $B_1$ and $B_2$ in the Landau expansion

$$F = A(P_1^2 + P_2^2 + P_3^2) + \frac{B_1}{4!}(P_1^4 + P_2^4 + P_3^4)$$

$$+ \frac{B_2}{4}(P_1^2P_2^2 + P_2^2P_3^2 + P_3^2P_1^2) + \frac{\Gamma_1}{6!}(P_1^6 + P_2^6 + P_3^6)$$

$$+ \frac{\Gamma_2}{48}[P_1^2(P_2^4 + P_3^4) + P_2^2(P_3^4 + P_1^4) + P_3^2(P_1^4 + P_2^4)]$$

$$+ \frac{\Gamma_3}{8}P_1^2P_2^2P_3^2 + \frac{\Delta_1}{8!}(P_1^8 + P_2^8 + P_3^8) \quad (2)$$

$$+ \frac{\Delta_2}{6!2!}[P_1^6(P_2^2 + P_3^2) + P_2^6(P_3^2 + P_1^2) + P_3^6(P_1^2 + P_2^2)]$$

$$+ \frac{\Delta_3}{(4!)^2}(P_1^4P_2^4 + P_2^4P_3^4 + P_3^4P_1^4)$$

$$+ \frac{\Delta_4}{96}(P_1^4P_2^2P_3^2 + P_1^2P_2^4P_3^2 + P_1^2P_2^2P_3^4).$$

In the first order of the perturbation theory, the fluctuation contributions to these vertices are represented by the diagrams depicted in the figure.

These diagrams can be calculated by knowing the relationship for the correlation function (propagator) $G_{\alpha\beta}(\mathbf{q})$. In ferroelectric materials, the propagator $G_{\alpha\beta}(\mathbf{q})$ is off-diagonal with respect to the Cartesian indices $\alpha$ and $\beta$ due to the dipole forces and, moreover, exhibits a considerable crystalline anisotropy for the $BaTiO_3$ compound [5]. It is clear that the diagrams should be calculated with allowance made for both these circumstances. This means that it is necessary to invoke the most general expression for the propagator $G_{\alpha\beta}(\mathbf{q})$ allowed by cubic symmetry. This expression is known [19, 20]; however, it is very cumbersome and, hence, is of little use for specific calculations.

The problem can be simplified taking into account that, in our case, the dipole–dipole interaction is very strong. The measure of this interaction is the dipole gap $\Omega_{dip}$ in the spectrum of the critical branch, i.e., the difference between the frequencies of the longitudinal and transverse (soft) optical modes at $q \longrightarrow 0$. In ferroelectric materials, this quantity has the same order of magnitude as the frequencies of conventional ("hard") optical phonons. This allows us to ignore the longitudinal component of the correlator and to use the limit

$\Omega_{dip} \longrightarrow \infty$ as the reliable working approximation. In this limit [20], we have

$$G_{\alpha\beta}(\mathbf{q}) = \frac{k_B T}{\epsilon^{-1} + s^2 q^2 + f s^2 q_\alpha^2} \quad (3)$$

$$\times \left[ \Delta_{\alpha\beta} - \frac{q_\alpha q_\beta}{s^{-2}\epsilon^{-1} + q^2 + f q_B^2} \left( \sum_{\gamma=1}^{3} \frac{q_\gamma^2}{s^{-2}\epsilon^{-1} + q^2 + f q_\gamma^2} \right)^{-1} \right].$$

Here, $\epsilon = \epsilon_0 C/(T - T_c)$ is the permittivity; $\epsilon_0$ is the permittivity of free space; $C$ is the Curie constant; and the parameters $s$ and $f$ characterize the dispersion and the crystalline anisotropy of the soft mode spectrum, respectively. As it must be, the correlator given by formula (3) does not depend on the dipole gap and is purely transversal: $q_\alpha G_{\alpha\beta}(\mathbf{q}) = 0$.

In the calculation of the diagrams, it is useful to keep in mind that integrals of the type

$$\int G_{\alpha\beta}(\mathbf{q}) d\mathbf{q} \quad (4)$$

for the off-diagonal correlator components are identically equal to zero. This can be most easily checked using symmetry considerations. The point is that the graphs containing the propagator $G_{\alpha\beta}(\mathbf{q})$ with $\alpha \neq \beta$ should give rise to terms with odd powers of the projections of the polarization vector in the Landau expansion. Since the inclusion of fluctuations in any case cannot reduce the symmetry of the system, all contributions of this type disappear.

Thus, by determining the combinatorial and tensor factors that correspond to the diagrams in the figure, it is easy to obtain the relationships for the first-order corrections to the coefficients $B_1$ and $B_2$. These relationships have the form

$$\delta B_1^{(1)} = I\left(\frac{\gamma_1}{2} + \gamma_2\right) k_B T$$
$$+ I^2\left(\frac{\delta_1}{8} + \frac{\delta_2}{2} + \frac{\delta_3}{4} + \frac{\delta_4}{4}\right)(k_B T)^2, \quad (5)$$

$$\delta B_2^{(1)} = I\left(\gamma_2 + \frac{\gamma_3}{2}\right) k_B T$$
$$+ I^2\left(\frac{\delta_2}{4} + \frac{\delta_3}{4} + \frac{5\delta_4}{8}\right)(k_B T)^2, \quad (6)$$

where

$$I = \frac{1}{(2\pi)^3 s^2} \int G_{11}(\mathbf{q}) d\mathbf{q}. \quad (7)$$

In terms of the field theory, integral (7) is characterized by the ultraviolet divergence. However, over the entire range of temperatures under consideration, the relative permittivity of the barium titanate satisfies the condition $\tilde{\epsilon} = C/(T - T_c) \gg 1$; hence, we have $\epsilon^{-1} \ll s q_D$, where $q_D$ is the Debye momentum. This enables us to calculate integral (7) in the limit $\epsilon^{-1} \longrightarrow 0$ without loss of accuracy. In this limit, integral (7) does not depend on the temperature but is a rather complex function of the anisotropy parameter $f$. This function cannot be found analytically; however, when the arguments are not very large, it is well approximated by a segment of the corresponding Taylor series; that is,

$$I \cong \frac{q_D}{3\pi^2 s^2}\left(1 - \frac{1}{5}f + \frac{1}{15}f^2 - \frac{127}{5005}f^3\right). \quad (8)$$

In particular, on the interval $-1 < f < 1$ (physically, it is the most interesting interval), the difference between the approximate expression (8) and its exact analog does not exceed 1.4%.

Now, we compare our relationships derived for the fluctuation components of the coefficients $B_1$ and $B_2$ with the experimental results. In order to make particular inferences, it is necessary to have information on the constants $\gamma_i$ and $\delta_j$ in Hamiltonian (1). These data can be obtained from microscopic (quantum-chemical, first-principles, etc.) calculations; however, their current accuracy is rather low. In this respect, we will use experimental estimates. Specifically, we assume that the constants $\gamma_i$ and $\delta_j$ are close in magnitude to the corresponding coefficients in the Landau expansion (2). The numerical values of these coefficients, including the eighth-order coefficients, were recently determined by processing the experimental data on the permittivity and the spontaneous polarization [10, 12]. Wang et al. [12] obtained the following results:

$$\Gamma_1 = 1.0 \times 10^{12} \text{ Vm}^2 \text{C}^{-5},$$
$$\Gamma_2 = -1.06 \times 10^{11} \text{ Vm}^9 \text{C}^{-5},$$
$$\Gamma_3 = 4.41 \times 10^{11} \text{ Vm}^9 \text{C}^{-5},$$
$$\Delta_1 = 1.95 \times 10^{15} \text{ Vm}^{13} \text{C}^{-7}, \quad (9)$$
$$\Delta_2 = 3.64 \times 10^{14} \text{ Vm}^{13} \text{C}^{-7},$$
$$\Delta_3 = 1.61 \times 10^{14} \text{ Vm}^{13} \text{C}^{-7},$$
$$\Delta_4 = 0.898 \times 10^{13} \text{ Vm}^{13} \text{C}^{-7}.$$

Here, the value of the coefficient $\Gamma_1$, which depends on the temperature, is taken at the phase transition point. Substitution of results (9) as the values of the anharmonicity constants $\gamma_i$ and $\delta_j$ into relationships (5) and (6) gives

$$\delta B_1^{(1)} = 3.94 \times 10^{11} I k_B T + 4.69 \times 10^{14} I^2 (k_B T)^2, \quad (10)$$

$$\delta B_2^{(1)} = 1.14 \times 10^{11} I k_B T + 1.37 \times 10^{14} I^2 (k_B T)^2. \quad (11)$$

The integral $I$ in these formulas depends on the model parameters $s$ and $q_D$, so that its direct estimation is com-plicated. However, even in the absence of information on the numerical value of the integral $I$, relationships (10) and (11) allow us to compare the theoretical and experimental data.

According to the experimental data obtained in [12], the coefficients $B_1$ and $B_2$ in the Landau expansion are as follows:

$$B_1 = -4.39 \times 10^{10} + 9.60 \times 10^7 T \ (\text{VmC}^{-1}), \quad (12)$$

$$B_2 = -0.896 \times 10^{10} + 2.68 \times 10^7 T \ (\text{VmC}^{-1}). \quad (13)$$

It can be seen that both coefficients increase monotonically with an increase in the temperature, which agrees qualitatively with the theoretical predictions (see relationships (10), (11)). In this case, the difference in the form of the temperature dependences (quadratic, linear) is of little importance. The point is that the coefficients $B_1$ and $B_2$ were determined using the experimental results obtained in a narrow temperature range (20–40 K) near $T_0 \approx 400$ K, in which the parabola is well approximated by a linear function. Furthermore, the algorithm for processing the experimental data in [12] was aimed at revealing the simplest (i.e., linear) dependence of the coefficients $B_1$ and $B_2$ on the temperature $T$.

One more fact is also of particular interest. Although the magnitudes of the fluctuation contributions to the coefficients $B_1$ and $B_2$ cannot be determined because the integral $I$ is unknown, their ratio can be evaluated from relationships (10) and (11). It can be easily seen that this ratio is equal to 3.46 for the components $\delta B_1^{(1)}$ and $\delta B_2^{(1)}$ linear in the temperature $T$ (i.e., dependent on the anharmonic constants $\gamma_i$) and 3.42 for the corresponding quadratic terms. It is evident that the quantity $\delta B_1^{(1)}/\delta B_2^{(1)}$ should lie between the above values. Let us compare the obtained estimate with the corresponding experimental value. From relationships (12) and (13), we have

$$\left(\frac{\delta B_1^{(1)}}{\delta B_2^{(1)}}\right)_{\text{experim}} = 3.58. \quad (14)$$

Thus, the theoretical and experimental values of $\delta B_1^{(1)}/\delta B_2^{(1)}$ differ from each other by no less than 5%.

The question arises as to whether the situation changes significantly when the results of other experiments are taken into account. To the best of our knowledge, there is only one alternative set of experimental data on the coefficients in the Landau expansion, including the eighth-order terms. This set is as follows [10]:

$$\Gamma_1 = 9.32 \times 10^{11} \ \text{Vm}^9\text{C}^{-5},$$

$$\Gamma_2 = -9.36 \times 10^{10} \ \text{Vm}^9\text{C}^{-5},$$

$$\Gamma_3 = -2.00 \times 10^{10} \ \text{Vm}^9\text{C}^{-5}.$$

$$\Delta_1 = 1.56 \times 10^{15} \ \text{Vm}^{13}\text{C}^{-7}, \quad (15)$$

$$\Delta_2 = 3.64 \times 10^{13} \ \text{Vm}^{13}\text{C}^{-7},$$

$$\Delta_3 = 9.43 \times 10^{12} \ \text{Vm}^{13}\text{C}^{-7},$$

$$\Delta_4 = 1.31 \times 10^{12} \ \text{Vm}^{13}\text{C}^{-7}.$$

It can be seen that these values differ substantially from those obtained in [12]. This fact is not surprising because these data were obtained under the assumption that all anharmonicity coefficients do not depend on the temperature. The expressions obtained for the fluctuation contributions to the coefficients $B_1$ and $B_2$ with the use of data (15) have the form

$$\delta B_1^{(1)} = 3.72 \times 10^{11} I k_B T + 2.16 \times 10^{14} I^2 (k_B T)^2, \quad (16)$$

$$\delta B_2^{(1)} = -1.04 \times 10^{11} I k_B T + 0.123 \times 10^{14} I^2 (k_B T)^2. \quad (17)$$

The quantity $\delta B_2^{(1)}$ in this case is not a monotonic function of the temperature. This fact does not directly contradict the experimental relationship (13), because formula (17) can describe the increase in the quantity $\delta B_2^{(1)}$ with an increase in the temperature $T$ in the high-temperature range. In the given case, another circumstance appears to be very important. As can be seen from the above example, the fluctuation corrections to the coefficient $B_2$ due to the sixth-order and eighth-order anharmonicities can have different signs and, hence, partially compensate each other. This means that there exists a mechanism responsible for the decrease in the quantity $\delta B_2^{(1)}$, which thus contributes to the smallness of this correction as compared to the correction $\delta B_1^{(1)}$. It is not ruled out that the above mechanism is responsible for the absence of the temperature dependence of the coefficient $B_2$ reconstructed from the experimental data in [4, 6, 7], while a pronounced temperature dependence of the coefficient $B_1$ was revealed in these papers.

In conclusion, we note that the effect studied in the present work has an analog in the dynamic theory of conventional weakly anharmonic lattices. The case in point is the temperature dependence of the elastic moduli of the crystal. This dependence arises from the anharmonic renormalization of the elastic moduli; i.e., it is explained by the presence of the corrections that change with a variation in the temperature. The difference between our case and the conventional case is that we consider the fluctuation renormalization of the lowest-order anharmonic coupling constants rather than the harmonic coupling constants.

ACKNOWLEDGMENTS

The author would like to thank A.S. Salasyuk for performing some of the control calculations.

This study was supported by the Russian Foundation for Basic Research (project no. 07-02-00345).

*Translated by O. Borovik-Romanova*